\documentclass[12 pt]{article}
\usepackage{graphicx}
\usepackage{color}
\begin{document}
\title{\bf Simulating the JUNO Neutrino Detectors}
\author{\bf Srikanta Sinha\\
            3A, Sharda Royale Apt.,\\
            G. M.Palya, Bengaluru-560 075, INDIA.\\
	    email: sinha.srikanta@gmail.com}
\maketitle

\newpage

\section{\bf ABSTRACT}
   The JUNO Neutrino detector system is simulated
using Monte-Carlo and analytical methods. 
A large numer of proton decay events are also
simulated.
Preliminary results from this endeavour are 
presented in the present article.

\section{\bf INTRODUCTION}
   The existence of neutrinos was first postulated by Wolfgang Pauli [1]
in order to explain the apparent violation of fundamental conservation laws in the
phenomenon of nuclear beta decay. Still for a long time experimental
searches failed to detect these highly elusive particles. This was
because neutrinos interact with matter neither through the familiar
electro-magnetic interactions nor through the nuclear strong interactions.
Neutrinos interact with matter primarily through the nuclear weak
interaction for which the interaction cross section at laboratory energies
are extremely small, typically, say, of the order of $10.0^{-44}$ $cm{-2}$.

   However, there existed on this Earth (once upon a time- may their
tribe increase!) 
daring and stubborn 
experimental physicists like Cowan and Reines [2] who took up the ardent
task to build a completely unconventional detector that consisted of
tons of an organic liquid scintillation detector placed it underground.
It is crucial for this type of experiments to keep the detector very
deep underground (typically, say, at a depth of $1 km$ or more- more is
better) since the Earth's surface is incessantly bombarded by high energy
secondary particles that result from the interactions of high energy
primary cosmic ray particles (mainly protons, alpha particles and
numerous other species of heavier nuclei) that originate
somewhere in the Universe and strike the Earth's atmosphere from
outside. Majority of the secondary particles produced are either
electrons or positrons or a few mesons or hadrons which are rather
easily filtered out by the Earth's surface. But there are again some
utterly useless (only for this type of experiments, they are truly
very interesting and the subjects of studies in other type of
experiments) ones which penetrate the Earth's surface and are 
detectable in sufficiently large (relatively speaking) numbers even
in laboratories located deep underground. This is the source 
of much trouble and hence one
has to have sufficient strength and courage to take up this type of
investigations. However, mother Nature always favours Her children
who are brave, honest and hard-working. Therefore, Reines and Cowan became
the first two to discover the fact that neutrinos are not the objects
of fantasy of mad people but they really exist in this Universe. Thus the
neutrino show began.

      Though we are able to see (i.e. detect) only a few neutrino events
say, in a year of operation,
even with large (say, about a few hundred tons) detectors, their actual
flux is very high (on the Earth's surface neutrinos that arrive do not have
the same origin- some of them are produced in the core of the mighty Sun-
in the thermonuclear fusion reactions that fuse hydrogen into helium and
produce energy, in Supernovae explosions which are the cataclysmic explosions 
that mark the deaths of massive stars, and a host of other astrophysical 
processes). Large fluxes of neutrinos are produced in nuclear fission 
reactors that are used by many countries for power generation. In high
energy particle accelerators such as the LHC, high energy neutrinos
are produced when short-lived particles (there are a very large 
number of types of such particles) decay.

     First confirmative observation of the phenomenon of
neutrino oscillations came from the Japanese Super-Kamiokade 
(SUPER-K) experiment that operates in the Kamioka mines.

     The stage is set now for new generation experiments and the JUNO
experiment is one of the very important and promising, i.e. we hope
that this experiment has the much needed high fiducial mass, very good
energy resolution and other parameters which will lead to the solution
of the fundamental questions being raised in this field.     

     So, at the present moment of GREAT IGNORANCE, what are the burning 
questions of neutrino physics ?

   Question 1: What is the neutrino mass hierarchy ?
               Let us try to be slightly more explicit. Neutrinos come in
three different flavours (i.e. there are three (so far observed) distinct
species of neutrinos- (i) the electron neutrino ($\nu_{e}$), (ii) the
muon neutrino ($\nu_{mu}$), and (iii) the tau neutrino ($\nu_{tau}$).
The important thing is we do not have answer to this question: which of these
neutrinos' mass is the least, which of these three types is the heaviest
and which neutrino species' mass lies midway between the other two.

   Question 2: What are the precise values of the neutrino oscillation
parameters ?

               To understand this question correctly, first we have to 
understand what is meant by neutrino oscillations ?

               As we said earlier neutrinos come in three distinct flavours
or types. Now, if we set up an experiment, say, to study a partcular type
of neutrino, say, the electron type neutrino, ($\nu_{e}$), we end up 
seeing somewhat less number of neutrino events than what we expect to
see. Now, this seems to be a paradox. It appears that some of these
neutrinos have simply disappeared from the scene ! There is a belief
among ordinary people (and physicists also seem to share this belief
that something which truly existed cannot simply vanish or into
non-existence). Something that have disappered must reappear in some
other form, may be in a different space-time point. Truly, this belief 
was reinforced again by the discovery of the phenomenon of neutrino
oscillations. If an electron neutrino ($\nu_{e}$) disapperas it reappears
again as a different neutrino species- either as a muon neutrino
($\nu_{mu}$) or as a tao neutrino ($\nu_{tau}$). But there is one important
thing to remeber: the probability (the fraction of times) of an electron
neutrino ($\nu_{e}$) transforming into a muon neutrino ($\nu_{mu}$) is
not the same of that transforming into a tao neutrino ($\nu_{tau}$).
This probability again is a function of (i.e. dependent upon) other
physical parameters, like energy of the neutrino etc.

Extremely large volumes (several tens of meters in diameter and height)
and massive (few kilotons or even a few tens of kilotons) are needed to
have meaningfully large number of detections per year. 

   Water Cerenkov detectors or Liquid Scintillator based detector systems
are generally used for this type of experiments. 
  The mean energies of neutrinos and their spectral energy distributions produced in 
different natural and man-made processes are, however, quite different.

    The JUNO detector has been designed in such a way that the data obtained from this
experiment will help the physicists answer the above mentioned questions.
 
    In addition to the above mentioned topics, the JUNO detector will also
study (i) neutrinos produced in supernova explosions, (ii) atmospheric
and geo-neutrinos, and (iii) proton decay events.
\section{\bf THE DETECTOR SYSTEM}
   The JUNO (Jiangmen Underground Neutrino Observatory) experiment is
located at Kaiping, Jiangmen in Southern China [3]. The primary aim of this
experiment is to study reactor neutrinos. Hence it is located at a mean
distance of $53$ km from two Nuclear Power Plants (Yangshing 4 cores and
Taishin, 6 cores). To shield the detectors from the large flux of high
energy muons resulting from cosmic ray interactions in the Earth's
atmosphere, the laboratory is located under the Dashi hill (overburden 
of $700$ m rocks). 

    The details of the JUNO NEUTRINO DETECTOR SYSTEM is available elsewhere[2].

    The detector system consists of three distinct SUBSYSTEMs, viz. (i) the
CENTRAL detector, (ii) the VETO detector, and (iii) the MUON TRACKER system.

    The JUNO CENTRAL detector consists of a huge spherical chamber (having a wall
made of Acrylic and having a diameter equal to $35.4$ m). This spherical chamber is
filled with nearly $20$ kilotons of Linear Acrylic Benzene (LAB) based liquid
scintillator. The fluor material used is PPO (2.5-diphenyloxazole). The doping 
concentration is $3$ g/L of the liquid. An additional wavelength shifter 
material (p-bis-(omethylstyryl)-benzene) or (bis-MSB) is used. Its concentration is
$15$ mg/L. This large volume of the liquid scintillator is viewed by two DISTINCT
PMT (photo-multiplier) systems. The LARGE PMT (LPMT) system consists of $18,000$
PMTs (each having diameter equal to $20$ inches). The SMALL PMT (LPMT) system
consists of $25,000$ PMTs (each having diameter equal to $3$ inches). The combined
photo-cathode coverage is larger than $75\%$. The smaller PMTs have less dark 
noise and better temporal response compared to the larger ones.

     The energy resolution of the JUNO CENTRAL detector is expected to be about
$3\%$ at $1$ MeV.

     The Acrylic sphere containing the LAB based liquid scintillator is completely
enclosed within a larger CYLINDRICAL chamber filled with ultra-pure water (Fig.1).
This cylindrical chamber has a diameter equal to $43.5$ m and a height equal to
$44$ m. This water volume is viewd by $2,000$ additional large ($20$ inch diameter)
PMTs. This water volume serves as a Cerenkov VETO detector. It helps in 
eliminating MUON induced events and also acts as a SHIELD against the radiation
from the rocks.

     On top of the cylindrical chamber there is a MUON TRACKING DETECTOR.
\section{\bf SIMULATING THE DETECTORS and EVENTS}
    Some details about the simulation methods and procedures 
for the particle and photon interactions (electro-magnetic processes,
such as the photo-electric interaction, coherent and incoherent
scattering of photons, pair production etc.) are available
elsewhere [4], [5].

    The detector geometrical configuration (the inner SPHERICAL detector 
and the outer CYLINDRICAL VETO detector)
including the position of each PMT are
calculated using standard analytical formulae. 
   Since in the present case we are dealing with high energy phenomena, we
have to use Monte Carlo codes to simulate high energy photon and particle
interactions that ultimately produce electro-magnetic cascades in the
detection medium. For this purpose we use the simulation procedures
described in [4], [5] and the MC code developed originally
by Vatcha [6]. However, we have restructured and significatly modified
the original code and use this for the present simulations instead of
any standard code, eg. GEANT4 etc. The photon interaction cross-sections
for different processes are calculated using the XCOM program developed
by Berger and Hubbell [7].

   As first trials we inject gamma rays (a total number of $6000$ events)
exactly at the center of the JUNO central detector and let them interact
within the detector volume. The scintillation photons are collected by the
PMTs (both the LPMTs and the SPMTs) and converted to a number of photo-
electrons. These photo-electrons are then multiplied in the dynode chains
within the photo-tubes and the resulting charges are converted into pulse
heights using a suitable capacitance.

\subsection{\bf Simulation of Proton Decay Events:}
  We have tried to simulate nucleon (in the present case only protons are
considered) decay events within the JUNO central detector. A total of
$2000$ such events are generated. These events are distributed within a
fiducial volume of $15$ meter radius (spherical volume). 
Only the ($e{+}-\pi_{0}$) decay mode of the proton is considered.
The resulting decay products are allowed to interact further within the
scintillation detector. The total number of scintillation photons are
very large in this case, since the proton rest mass is close to $1 GeV$.
We should mention that we consider only the decay of protons at rest.
Nuclear effects are not taken into account in the present calculations.

   Another important aspect of the new MC code is that it is written almost 
entirely using the FORTRAN 90/95
language.
\section{\bf RESULTS}
     A computer simulated image of the JUNO central detector is shown in Fig.1.
Since the number of PMTs is too large, all the individual PMTs are not resolved
in the figure. The large PMTs (each of them having a diameter of $20$ inches)
are shown in red color, while the small PMTs (each of them having a diameter
of $3$ inches) are shown in green color.
\begin{figure}
\centering
\includegraphics[width=9cm, height=8cm]{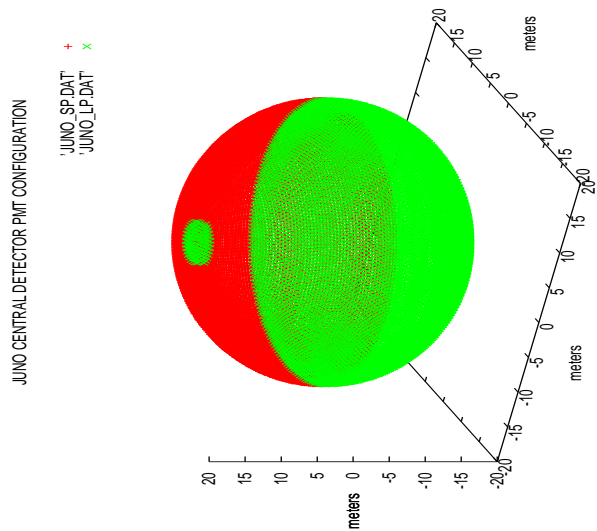}
\caption{The JUNO CENTRAL DETECTOR}
\end{figure}
   The simulated image of the JUNO central detector enclosed within the
JUNO cylindrical VETO detector is shown in Fig.2.
\begin{figure}
\centering
\includegraphics[width=9cm, height=8cm]{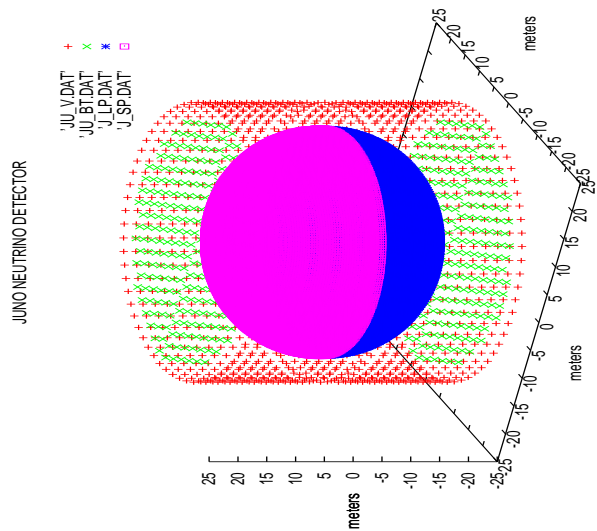}
\caption{The JUNO CENTRAL DETECTOR enclosed within the VETO (CYLINDRICAL) DETECTOR}
\end{figure}
  We show the pulse height spectrum that results from $1 MeV$ gamma rays
interacting at the center of the JUNO central detector. The scintillation
photons that are collected by the PMTs are converted into photo-electrons.
These photo-electrons undergo secondary multiplication in the dynode
chain within the PMTs.
\begin{figure}
\centering
\includegraphics[width=9cm, height=8cm]{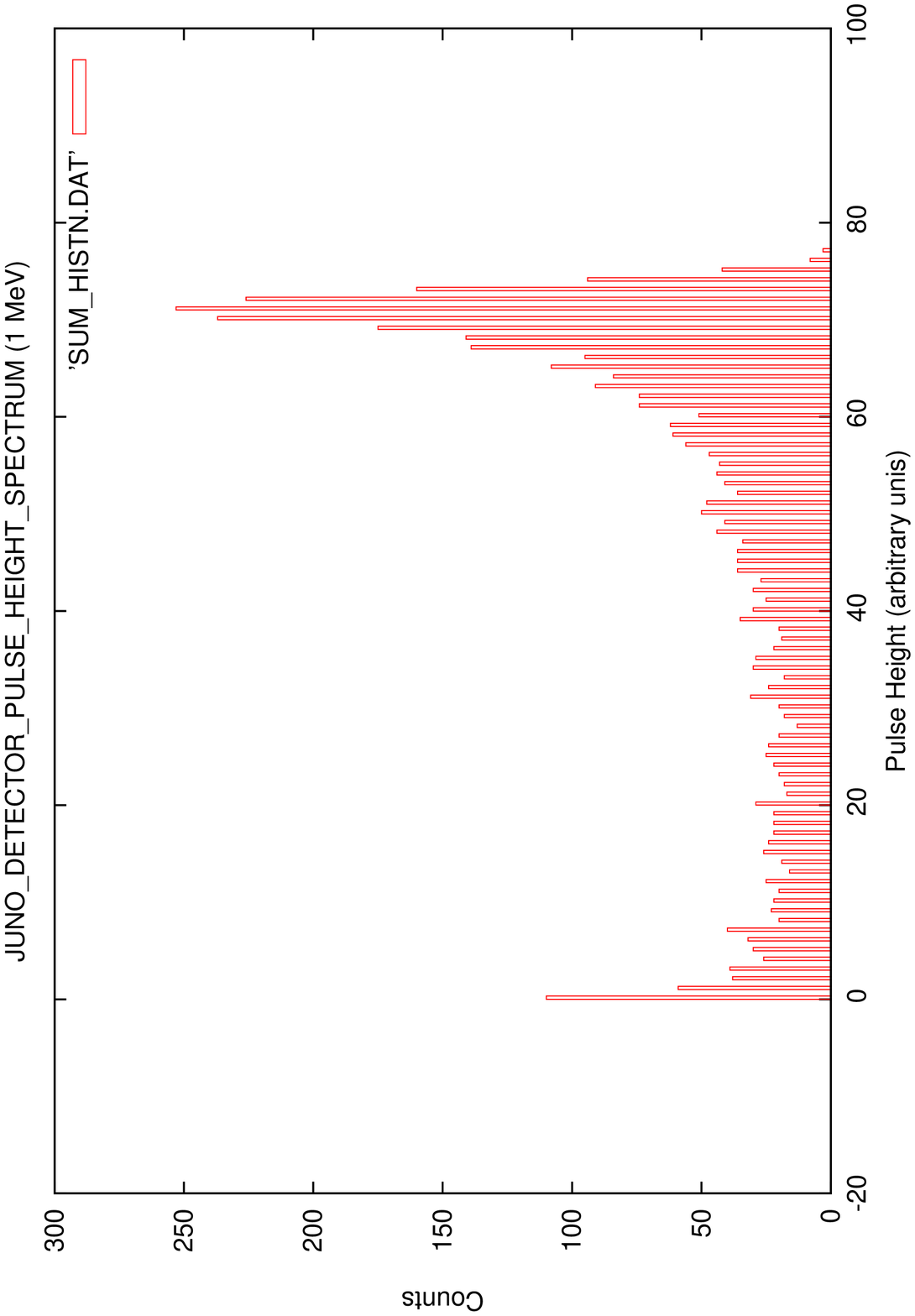}
\caption{The JUNO pulse height spectrum for photons (energy $1 MeV$) 
interacting within the CENTRAL DETECTOR}
\end{figure}
  We have made an attempt to simulate nucleon (in the present case
only protons) decay events within the JUNO central detector. Fig.4
shows the frequency distribution of the number of photo-electrons
that are seen by the central detector PMTs. 
\begin{figure}
\centering
\includegraphics[width=9cm, height=8cm]{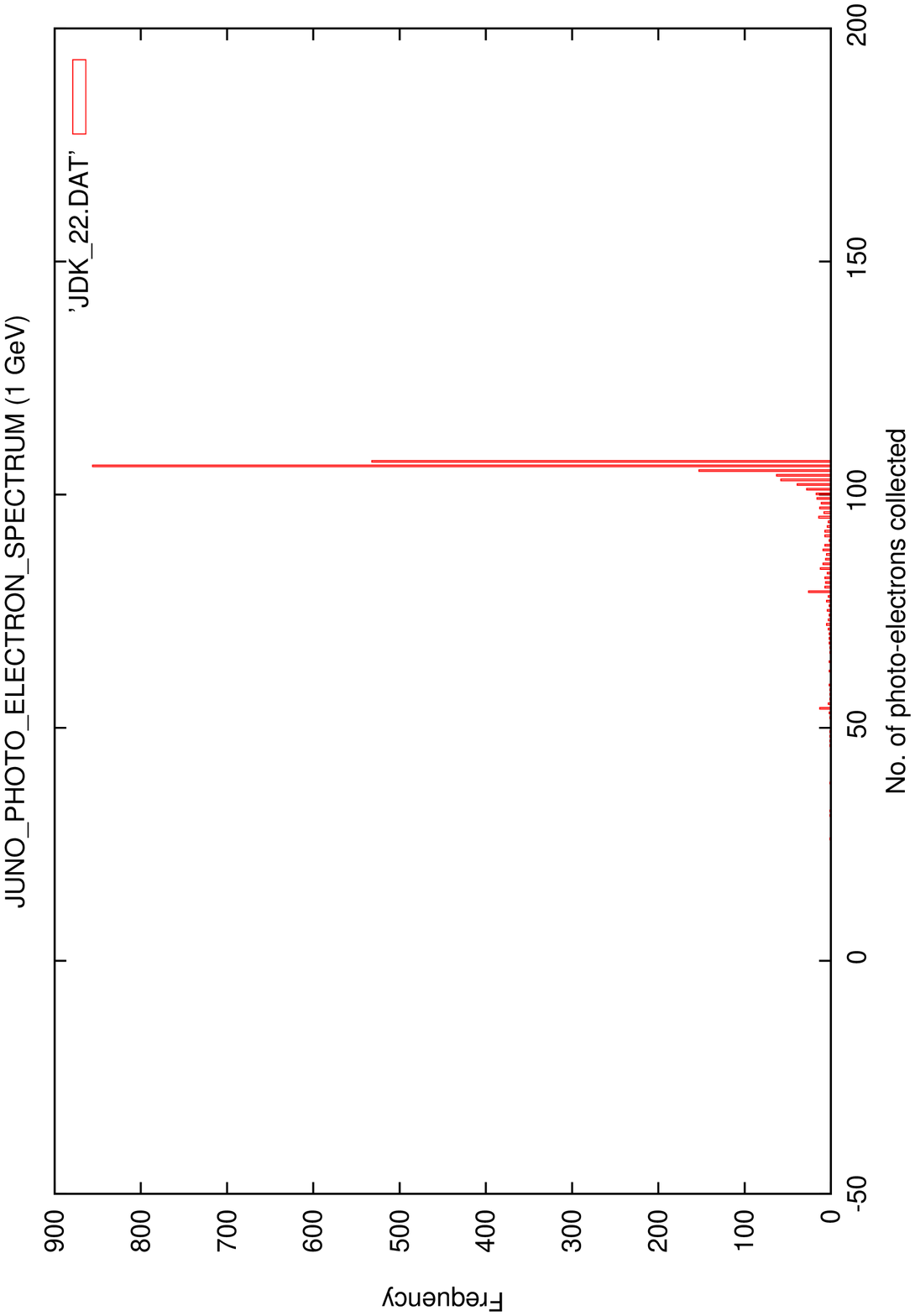}
\caption{The frequency distribution of the number of photo-
electrons in proton decay events. The peak corresponds to
almost $2.5$ million photo-electrons.}
\end{figure}
\section{\bf DISCUSSIONS and CONCLUSION}
    Detailed physics have to be incorporated to make the
simulations much more realistic. Some of these processes have
already been included in the latest procedures and test runs taken.
Careful checking is required. These will help in getting
accurate estimates of physical parameters. The simulation
procedures are to be used to create different event topologies. 

\section{\bf ACKNOWLEDGEMENTS}
    I would like to express my deep sense of gratitude to all my
teachers, especially to (Late) A.N. Ghosh, (Late) K. Sivapasad,
Prof. A.K. Biswas and Dr. B.K. Chatterjee.
They had been and still remain great sources of inspiration and support to me.
I am greatly indebted to (Late) Prof. John H. Hubbell of the NIST
for kindly providing me the XCOM software. All the photon
interaction cross-sections used in the present work are
calculated using this program. I also
thank the JUNO collaboration to make lot of information available
about this great experiment on their website.

\end{document}